%% file: jmorris.tex
\documentclass[10pt]{article}
\usepackage{graphicx}
\usepackage{subcaption}
\usepackage{placeins}
\usepackage{lineno}

\newcommand{\eVdist}{\kern-0.06667em}

\def\Title#1{\begin{center} {\Large #1 } \end{center}}
\def\Author#1{\begin{center}{ \sc #1} \end{center}}
\def\Address#1{\begin{center}{ \it #1} \end{center}}

\newcommand\pubblock{\rightline{\begin{tabular}{l} Proceedings of the Second Annual LHCP\\ \pubnumber\\
         \pubdate  \end{tabular}}}

\newenvironment{Abstract}{\begin{quotation} \begin{center} 
             \large ABSTRACT \end{center}\bigskip 
      \begin{center}\begin{large}}{\end{large}\end{center} \end{quotation}}

\newenvironment{Presented}{\begin{quotation} \begin{center} 
             PRESENTED AT\end{center}\bigskip 
      \begin{center}\begin{large}}{\end{large}\end{center} \end{quotation}}

\input econfmacros.tex

\usepackage{color}

\newcommand\pubnumber{ ATL-PHYS-PROC-2014-224 }

\newcommand\pubdate{\today}

\def\affiliation{
On behalf of the ATLAS Collaboration, \\
Queen Mary University of London \\
Mile End Road, E1 4NS, UK }

\begin{document}

\large
\begin{titlepage}
\pubblock

\vfill
\Title{ Top quark pair production cross section at LHC in ATLAS  }
\vfill

\Author{ John Morris  }
\Address{\affiliation}
\vfill
\begin{Abstract}
Measurements of the top quark production cross section in proton-proton collisions with the ATLAS detector at the Large Hadron Collider are presented. The measurements require no, one or two electrons or muons in the final state (single lepton, dilepton, hadronic channel).
In addition, the decay modes with tau leptons are tested (channels with tau leptons). The main focus is on measurements of differential spectra of $t\bar{t}$ final states, in particular, measurements that are able to constrain the modelling of additional parton radiation like the jet multiplicity distribution.

\end{Abstract}
\vfill

\begin{Presented}
The Second Annual Conference\\
 on Large Hadron Collider Physics \\
Columbia University, New York, U.S.A \\ 
June 2-7, 2014
\end{Presented}
\vfill
\end{titlepage}
\def\thefootnote{\fnsymbol{footnote}}
\setcounter{footnote}{0}
%

\normalsize 


\section{Introduction}
\vspace{-1.0em}
A summary of top quark pair production cross section measurements made with the ATLAS detector~\cite{Aad:2008zzm} at the LHC~\cite{Evans:2008zzb} is presented. Section~\ref{sec:diff7} presents the differential cross section measured in the $\ell$+jets channel at $\sqrt{s}=7$ TeV over a range of kinematic variables. This provides a test of different Monte Carlo generators, theoretical models and Parton Density Function (PDF) sets. Section~\ref{sec:emu8} presents an inclusive cross section result in the di-lepton $e\mu$ final state at $\sqrt{s}=8$ TeV which achieves, for this observable, the highest precision of any ATLAS result. Section~\ref{sec:ellJ8} presents an inclusive cross section result in the $\ell$+jets channel. Finally, in Section~\ref{sec:summary}, a summary of all ATLAS top pair cross sections results, along with global comparisons, is presented.

\section{Differential cross sections in $\ell$+jets at $\sqrt{s}=7$ TeV}
\label{sec:diff7}
\vspace{-1.0em}
The top quark pair production cross section, using the full $\mathcal{L}=4.6$ ${\rm fb}^{-1}$ integrated luminosity from the 2011 $\sqrt{s}=7$ TeV data set, has been measured differentially with respect to four variables.

\begin{itemize}\addtolength{\itemsep}{-.35\baselineskip}
\item The transverse momentum of a single top quark, $p_{\rm T}^{t}$, which is sensitive to higher order corrections and new physics signals in the high $p_{\rm T}$ tail.

\item The mass of the $t\bar{t}$ system, $m_{t\bar{t}}$, which is sensitive to possible deviation from the standard model due to exotic resonances.

\item The transverse momentum of the $t\bar{t}$ system, $p_{\rm T}^{t\bar{t}}$, which is sensitive to additional radiation.

\item The rapidity of the $t\bar{t}$ system, $y_{t\bar{t}}$, which provides important constraints to PDFs, especially at high x.
\end{itemize}
A summary of the main results of this measurement are presented, for more details please see \cite{Aad:2014zka}.

\subsection{Event selection}
\label{sec:evtSel}
\vspace{-0.5em}
Events are selected by requiring a single lepton trigger for either the electron or muon channel. All events are required to have passed a good runs list to ensure functional operation of all components of the detector. 

High quality isolated electrons are required to have $p_{\rm T}>25$ GeV and a pseudorapidity $|\eta|<2.47$ with a veto on electron candidates in the transition region between the barrel and end-cap calorimeters $\left(1.37 < |\eta| < 1.52\right)$. High quality isolated muons are required to have $p_{\rm T}>25$ GeV and a pseudorapidity $|\eta|<2.5$. At least four jets are required that have been reconstructed using the anti-$k_{t}$ algorithm \cite{Cacciari:2008gp} from topological clusters of energy depositions with a distance parameter of 0.4. Only jets with $p_{\rm T}>25$ GeV and $|\eta|<2.5$ are considered in the analysis. To select jets that come from the hard scattering and to suppress pile-up, the jet vertex fraction, defined as the sum of the $p_{\rm T}$ of tracks associated with the jet and originating from the primary vertex divided by the sum of the $p_{\rm T}$ from all tracks associated with the jet, is required to be greater that 0.75. At least one jet is required to be $b$-tagged at a 70\% efficient working point.

The missing transverse momentum $E_{\rm T}^{miss}$ is derived from the vector sum of calorimeter cell energies within $|\eta|<4.9$ and corrected on the basis of the dedicated calibrations of the associated physics objects. Significant cells not associated with high-$p_{\rm T}$ physics objects are included. Events must have \\ $E_{\rm T}^{miss}>35$ GeV and a $W$ transverse mass, $m_{\rm T}^{w}>35$ GeV\footnote{$m_{\rm T}^{w}=\sqrt{2 p_{\rm T}^{\ell}p_{\rm T}^{\nu}\left( 1 - \cos \left(\phi^{\ell} - \phi^{\nu}\right)\right)}$ where $p_{\rm T}^{\nu}$ is identified with $E_{\rm T}^{miss}$ and $\phi^{\nu}$ is the $\phi$ component of the $E_{\rm T}^{miss}$ 4-vector.} 

A kinematic likelihood fitter~\cite{Erdmann:2013rxa} is used to fully reconstruct the $t\bar{t}$ kinematics and events are selected with a $\log \mathcal{L} > -50$.

\subsection{Results}
\vspace{-0.5em}
Events are unfolded using a regularized SVD \cite{Hocker:1995kb} method, which is found to reduce large statistical fluctuations that can be introduced when directly inverting the migration matrix. The Asymmetric Iterative BLUE~\cite{Group:2008vx} method is used to combine the results measured in the electron and muon channels, where BLUE refers to the best linear unbiased estimator.

\begin{figure}[ht!]
    \centering
    \begin{subfigure}[t]{0.45\textwidth}
        \centering
        \includegraphics[width=0.95\textwidth]{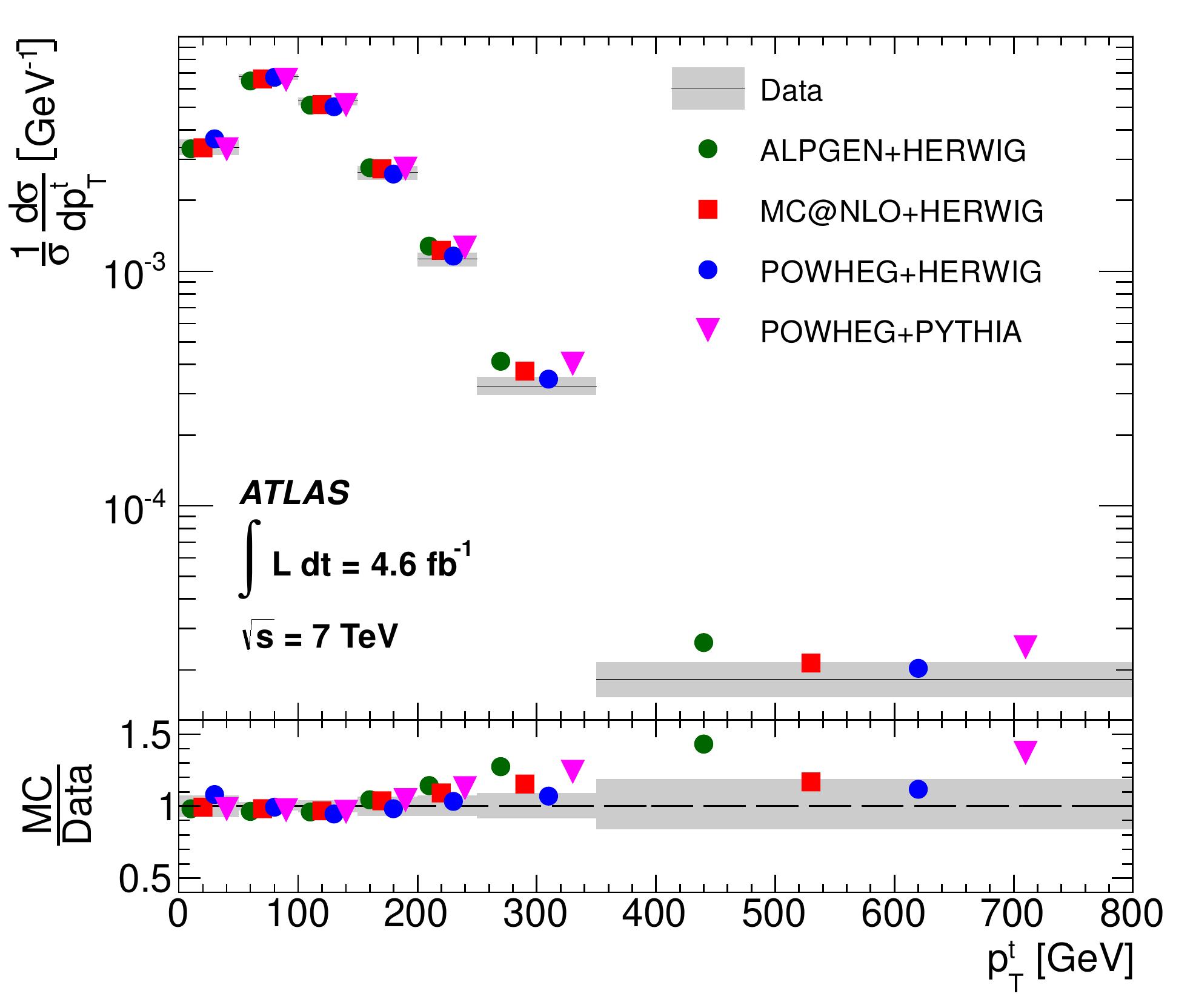}
        \caption{Transverse momentum of top quark}
    \end{subfigure}%
    ~ 
    \begin{subfigure}[t]{0.45\textwidth}
        \centering
        \includegraphics[width=0.95\textwidth]{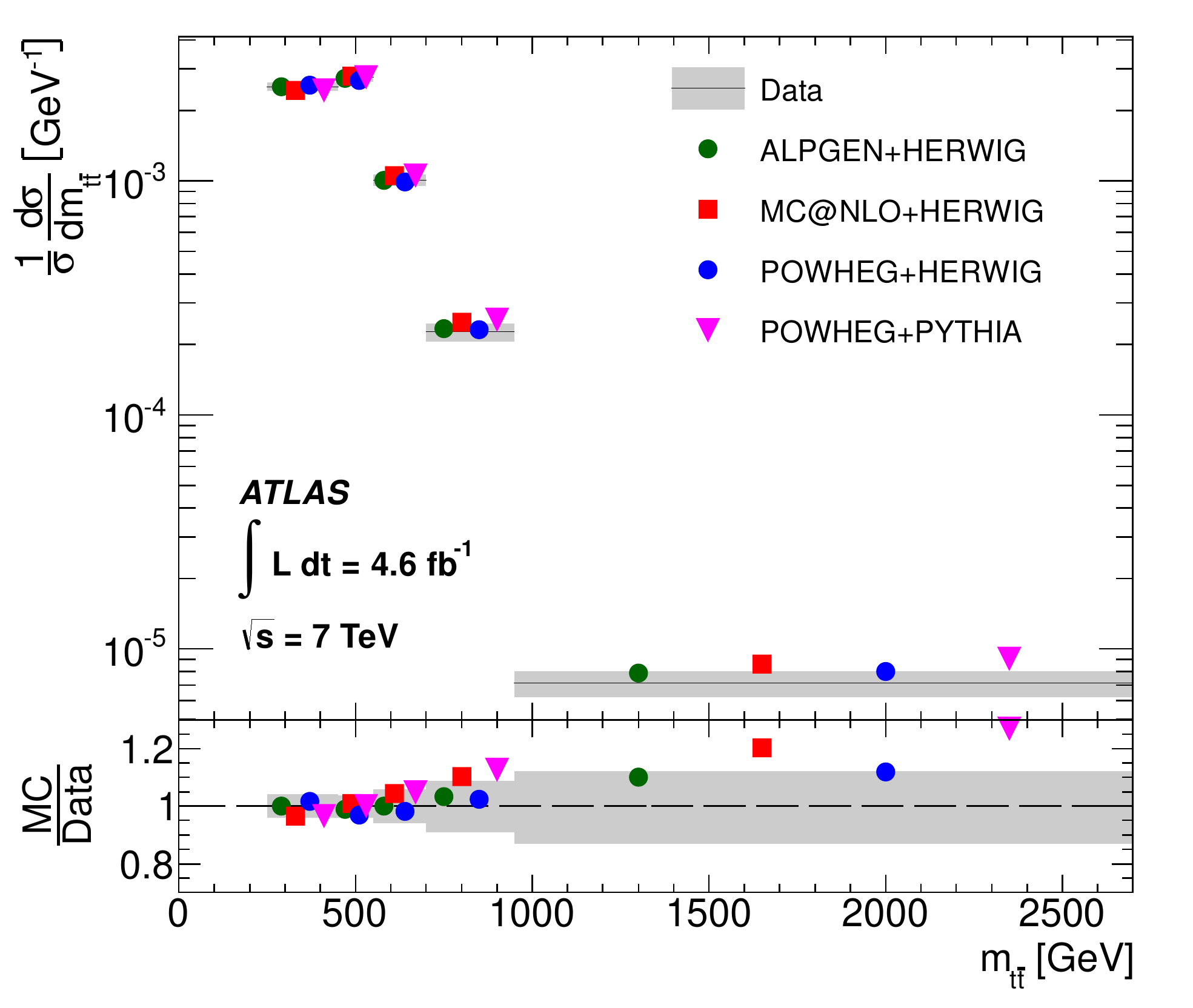}
        \caption{Mass of ${t\bar{t}}$ system}
    \end{subfigure}    
    
    \caption{Normalized differential cross sections. Comparisons to several generators are shown with points corresponding to {\sc Alpgen+Herwig} (circles), {\sc MC@NLO} (squares) and {\sc Powheg+Herwig} (triangles).}
    \label{fig:diff7_vsMCGen}
\end{figure}

Selected measured distributions, shown in Figure~\ref{fig:diff7_vsMCGen}, are compared with the predictions of different MC generators. All generators use {\sc Herwig}~\cite{Corcella:2000bw} for parton shower and hadronization, while the PDFs are different. The largest deviation between measurements and the predictions is observed in the case of $p_{\rm T}^{t}$ when comparing the measurement to {\sc Alpgen}~\cite{Mangano:2002ea} for large values of $p_{\rm T}^{t}$. In contrast, {\sc MC@NLO}~\cite{Frixione:2002ik} and {\sc Powheg}~\cite{Nason:2004rx,Frixione:2007vw} predict shapes closer to the measured distribution. There is a general trend observed in the data: the measured $p_{\rm T}^{t}$ is softer above 200 GeV than any of the predictions. On the other hand, all three MC generators describe the shape of $m_{t\bar{t}}$ reasonably well.

\begin{figure}[ht!]
    \centering
    \begin{subfigure}[t]{0.45\textwidth}
        \centering
        \includegraphics[width=0.95\textwidth]{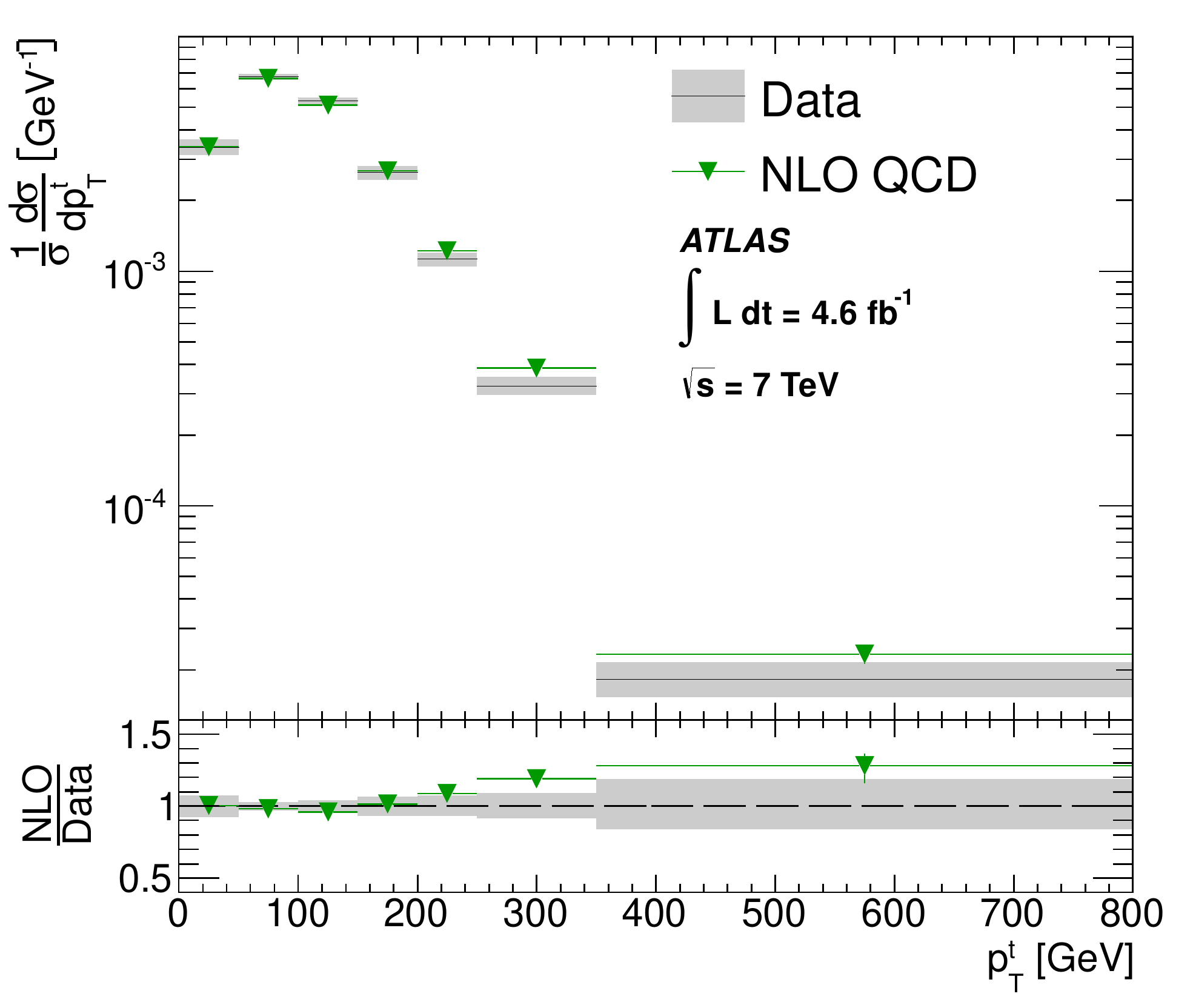}
        \caption{NLO QCD $p_{\rm T}^{t}$}
    \end{subfigure}%
    ~ 
    \begin{subfigure}[t]{0.45\textwidth}
        \centering
        \includegraphics[width=0.95\textwidth]{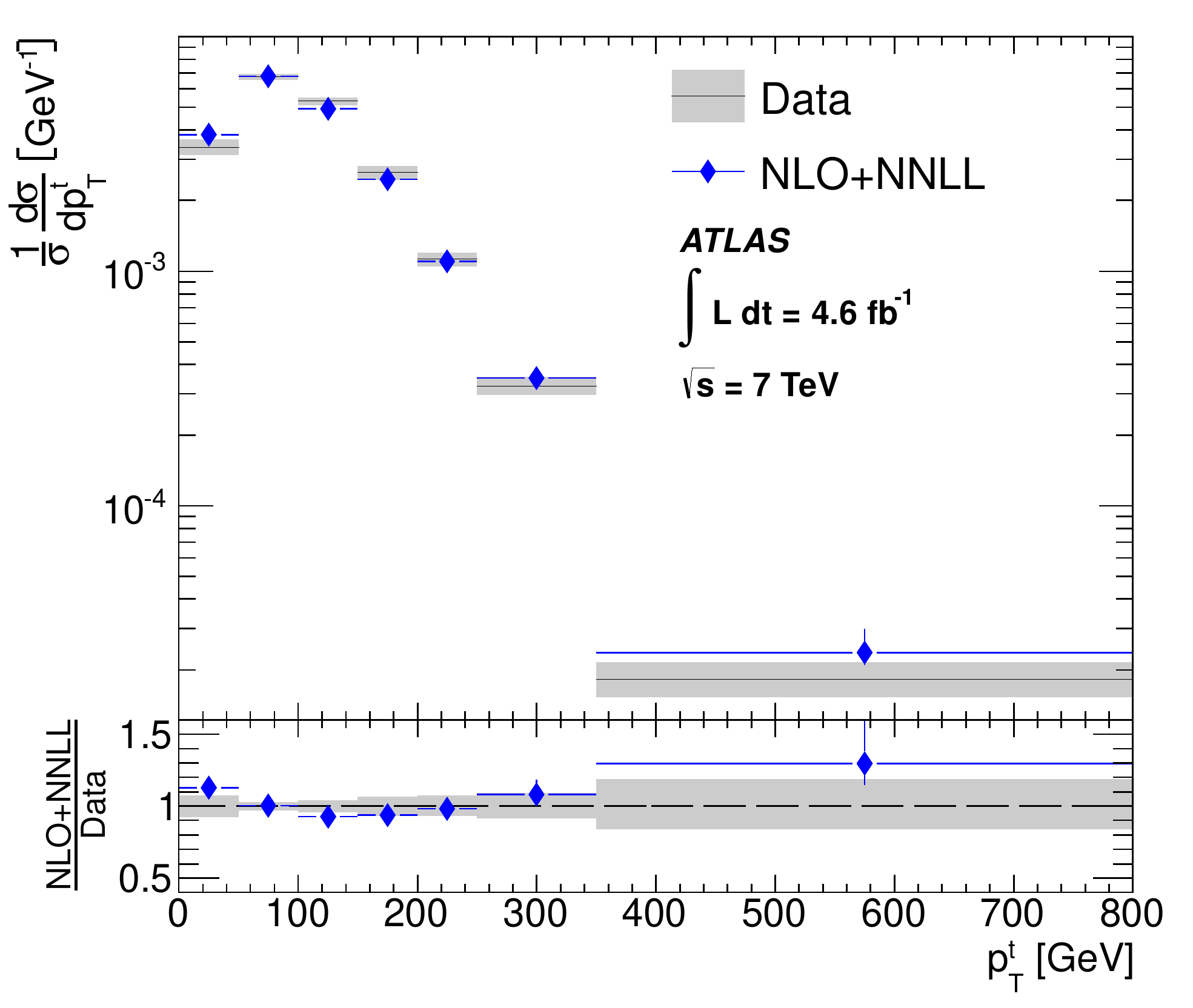}
        \caption{NLO+NNLL $p_{\rm T}^{t}$}
    \end{subfigure}    
    
    \caption{Normalized differential cross sections of $p_{\rm T}^{t}$ compared to (a) NLO QCD predictions based on MCFM with the CT10 PDF, and (b) NLO+NNLL calculations using the MSTW20008NLO PDF.}
    \label{fig:diff7_vsQCD}
\end{figure}

Selected distributions are also shown compared to QCD calculations at NLO (based on MCFM with the CT10~\cite{Lai:2010vv} PDF) and NLO+NNLL~\cite{Kidonakis:2010dk} (using the MSTW2008NNLO~\cite{Martin:2009iq} PDF) in Figure~\ref{fig:diff7_vsQCD}. These predictions do not include parton showering and hence cannot be precisely compared to the unfolded data. The uncertainties on the NLO predictions due to the PDFs have been evaluated at 68\% CL using the provided CT10 error-PDF set. Another source of uncertainty considered is the one related to the factorization and renormalization scales. The nominal value has been assumed to be $\mu=m_{t}=175.2$ GeV for both scales, and is varied up and down from $m_{t}/2$ to $2m_{t}$. For the NLO QCD prediction, the theory uncertainty is taken as the envelope of the scale variations with respect to the nominal added in quadrature with the PDF uncertainties. For the NLO+NNLL prediction, the uncertainty comes from the same fixed scale variations and, in addition in the case of $p_{\rm T}^{t}$, from the alternative dynamic scale $\mu=\sqrt{m_{t}^{2}+{p_{\rm T}^{t}}^{2}}$. The data is softer than both predictions in the tail of the $p_{\rm T}^{t}$ distribution, although not significantly.

\begin{figure}[ht!]
    \centering
    \begin{subfigure}[t]{0.45\textwidth}
        \centering
        \includegraphics[width=0.95\textwidth]{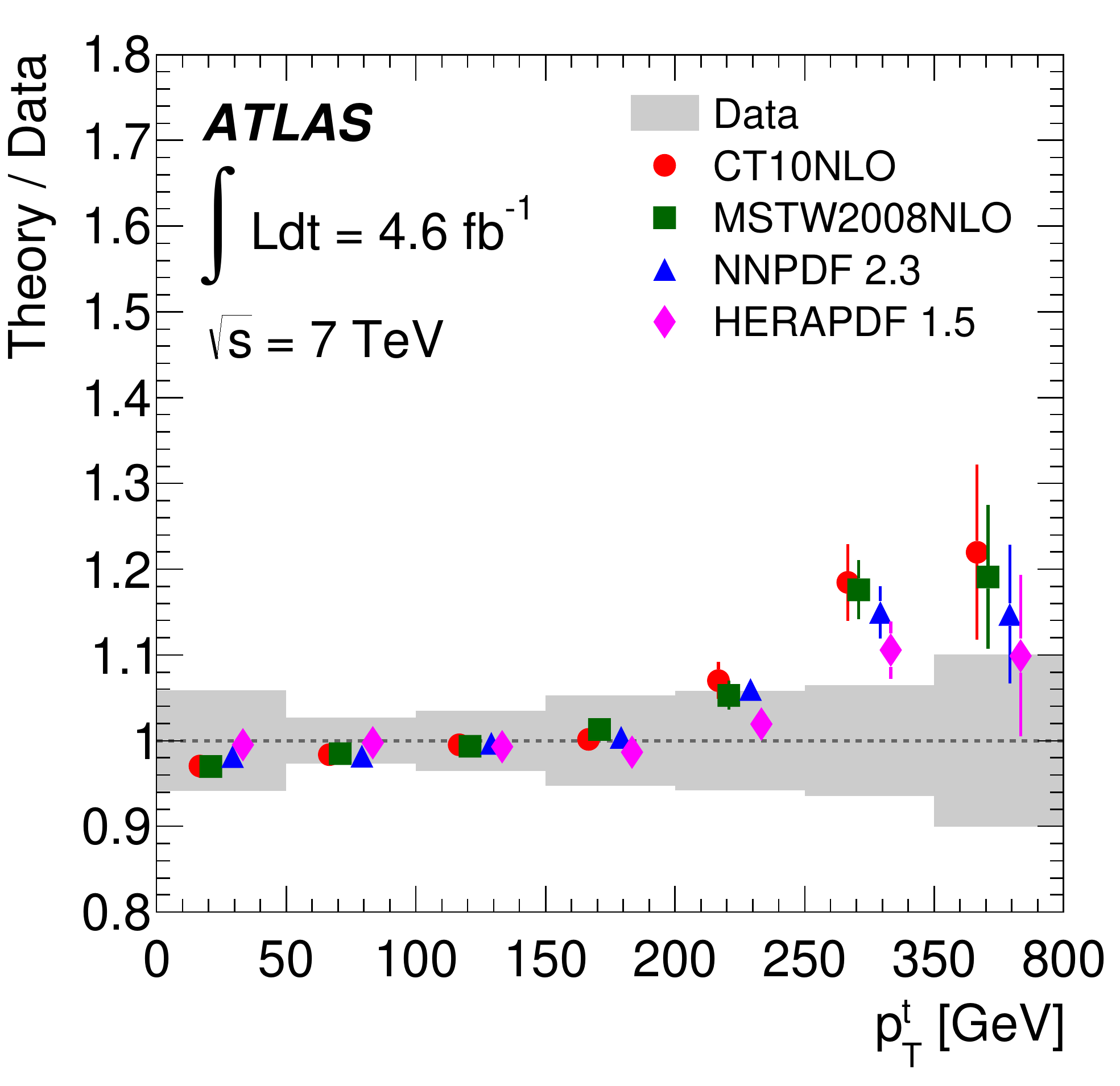}
        \caption{Transverse momentum of top quark}
    \end{subfigure}%
    ~ 
    \begin{subfigure}[t]{0.45\textwidth}
        \centering
        \includegraphics[width=0.95\textwidth]{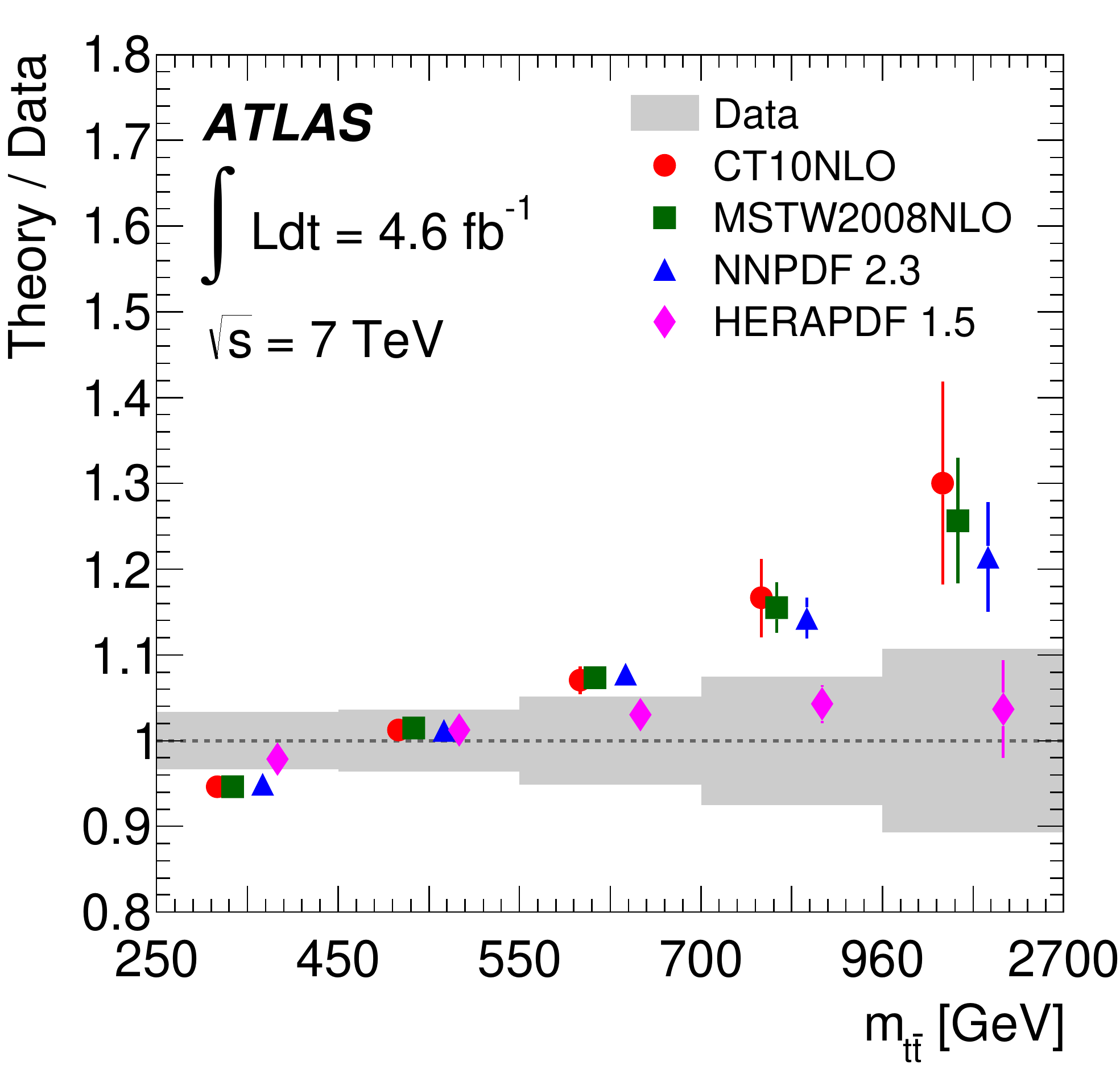}
        \caption{Mass of ${t\bar{t}}$ system}
    \end{subfigure}    
    
    \caption{Ratios of the NLO QCD predictions to the measured normalised differential cross-sections for different PDF sets (CT10, MSTW20008NLO, NNPDF2.3 and HERAPDF1.5).}
    \label{fig:diff7_vsPDF}
\end{figure}

The predictions of various NLO PDF sets are evaluated using MCFM version 6.5, interfaced to four different PDF sets: CT10, MSTW20008NLO, NNPDF2.3~\cite{Ball:2012cx} and HERAPDF1.5~\cite{Aaron:2009aa}. The uncertainties on the predictions include the PDF uncertainties and the fixed scale uncertainties already described. The comparisons between data and the different predictions are shown in Figure~\ref{fig:diff7_vsPDF} for the normalized differential cross-sections. A certain tension between data and all predictions is observed in the case of $p_{\rm T}^{t}$ at high values. For the $m_{t\bar{t}}$ distribution, one should note again that MCFM is effectively only a leading order calculation and resummation effects are expected to play an important role. 

\section{Inclusive cross section in $e\mu$ at $\sqrt{s}=8$ TeV}
\label{sec:emu8}
\vspace{-1.0em}
The top quark pair production cross section, using the full $\mathcal{L}=20.3$ ${\rm fb}^{-1}$ integrated luminosity from the 2012 $\sqrt{s}=8$ TeV data set, has been measured inclusively in the di-lepton $e\mu$ channel. This represents the most precise ATLAS top quark pair production cross section result~\cite{Aad:2014kva}. The event selection broadly follows what was outlined in Section~\ref{sec:evtSel}. Events must contain at least one $b$-tagged jet, with no kinematic likelihood selection applied. The $t\bar{t}$ production cross-section, $\sigma_{t\bar{t}}$, was determined by counting the number of opposite-sign $e\mu$ events with exactly one $\left(N_{1}\right)$ and exactly two $\left(N_{2}\right)$ $b$-tagged jets, ignoring any untagged jets which may be present. The two event counts can be expressed as:
\vspace{-0.5em}
\begin{eqnarray}
{\rm 1} \, b{\rm -tagged \, jet} : N_{1} &=& \mathcal{L}\sigma_{t\bar{t}}\epsilon_{e\mu}2\epsilon_{b}\left(1-C_{b}\epsilon_{b}\right) + N_{1}^{bkg},  \\
{\rm 2} \, b{\rm -tagged \, jets} : N_{2} &=& \mathcal{L}\sigma_{t\bar{t}}\epsilon_{e\mu}C_{b}\epsilon_{b}^{2} + N_{2}^{bkg},
\label{eq:richard2}
\end{eqnarray}

\vspace{-0.8em}
where $\mathcal{L}$ is the integrated luminosity of the sample and $\epsilon_{e\mu}$ the efficiency for a $t\bar{t}$ event to pass the opposite-sign $e\mu$ preselection. The combined probability for a jet from the quark $q$ in the $t \to Wq$ decay to fall within the acceptance of the detector, be reconstructed as a jet with transverse momentum above the selection threshold, and be tagged as a $b$-jet, is denoted by $\epsilon_{b}$. Although this quark is almost always a $b$ quark, $\epsilon_{b}$ thus also accounts for the approximately 0.2\% of top quarks that decay to $Ws$ or $Wd$ rather than $Wb$, slightly reducing the effective tagging efficiency. If the decays of the two top quarks and the subsequent reconstruction of the two $b$-tagged jets are completely independent, the probability to tag both $b$-jets, $\epsilon_{bb}$, is given by $\epsilon_{bb} =  \epsilon_{b}^{2}$. In practice, small correlations are present for both kinematic and instrumental reasons, and they are taken into account via the tagging correlation $C_{b}$, defined as $C_{b}=\epsilon_{bb}/\epsilon_{b}^{2}$. Backgrounds from sources other than $t\bar{t} \to e\mu \nu \bar{\nu} b \bar{b}$ also contribute to the event counts $N_{1}$ and $N_{2}$, and are given by the background terms $N_{1}^{bkg}$ and $N_{2}^{bkg}$. The preselection efficiency $\epsilon_{e\mu}$ and tagging correlation $C_{b}$ were taken from $t\bar{t}$ event simulation, and the background contributions $N_{1}^{bkg}$ and $N_{2}^{bkg}$ were estimated using a combination of simulation and data-based methods, allowing the two equations to be solved yielding $\sigma_{t\bar{t}}$ and $\epsilon_{b}$.

The dominant systematic uncertainties come from $t\bar{t}$ signal modelling, initial and final state radiation,  PDF uncertainties, lepton identification and jet energy scale and resolution.

The $pp \rightarrow t\bar{t}$ cross section is found to be:
\vspace{-0.8em}
\begin{equation}
\sigma_{t\bar{t}} = 242.4\pm1.7 \pm 5.5 \pm 7.5 \pm 4.2 \; {\rm pb}.
\end{equation}
where the four uncertainties arise from data statistics, experimental and theoretical systematic effecs, knowledge of the integrated luminosity and of the LHC beam energy.
\vspace{-1.0em}
\section{Inclusive cross section in $\ell$+Jets at $\sqrt{s}=8$ TeV}
\label{sec:ellJ8}
\vspace{-1.0em}
The top quark pair production cross section has been measured inclusively in the $\ell$+jets channel, using $\mathcal{L}=5.8$ ${\rm fb}^{-1}$ integrated luminosity from the 2012 $\sqrt{s}=8$ TeV data set. A summary of this measurement is presented, for more details please see \cite{ATLAS:2012jyc}.

The event selection broadly follows what was outlined in Section~\ref{sec:evtSel}. Electrons and muons are required to have a $p_{\rm T}>40$ GeV in order to reduce the $W$+jets and multijet background. The $e$+jets channel requires $E_{\rm T}^{\rm miss}>30$ GeV and $m_{\rm T}^{\rm W}>35$ GeV, while the $\mu$+jets channel requires $E_{\rm T}^{\rm miss}>20$ GeV and \\$E_{\rm T}^{\rm miss}+m_{\rm T}^{\rm W}>60$ GeV. At least 3 jets are required with one $b$-tagged jet at a 70\% efficient working point.

The number of $t\bar{t}$ events is obtained from the data using a likelihood discriminant template fit, constructed from the lepton pseudorapidity and the transformed event aplanarity.

The dominant systematic uncertainties come from $t\bar{t}$ signal modelling, initial and final state radiation,  PDF uncertainties, lepton identification and jet energy scale and resolution.

The $pp \rightarrow t\bar{t}$  cross section is found to be:
\vspace{-0.5em}
\begin{equation}
\sigma_{t\bar{t}} = 241\pm2 {\rm \left(stat\right)} \pm 31 {\rm \left(syst\right)} \pm 9 {\rm \left(lumi\right)} \; {\rm pb}.
\end{equation}
\vspace{-3.0em}
\section{Summary of all ATLAS top pair cross section results}
\label{sec:summary}
\begin{figure}[ht!]
    \centering
    ~ 
    \begin{subfigure}[t]{0.42\textwidth}
        \centering
        \includegraphics[width=0.95\textwidth]{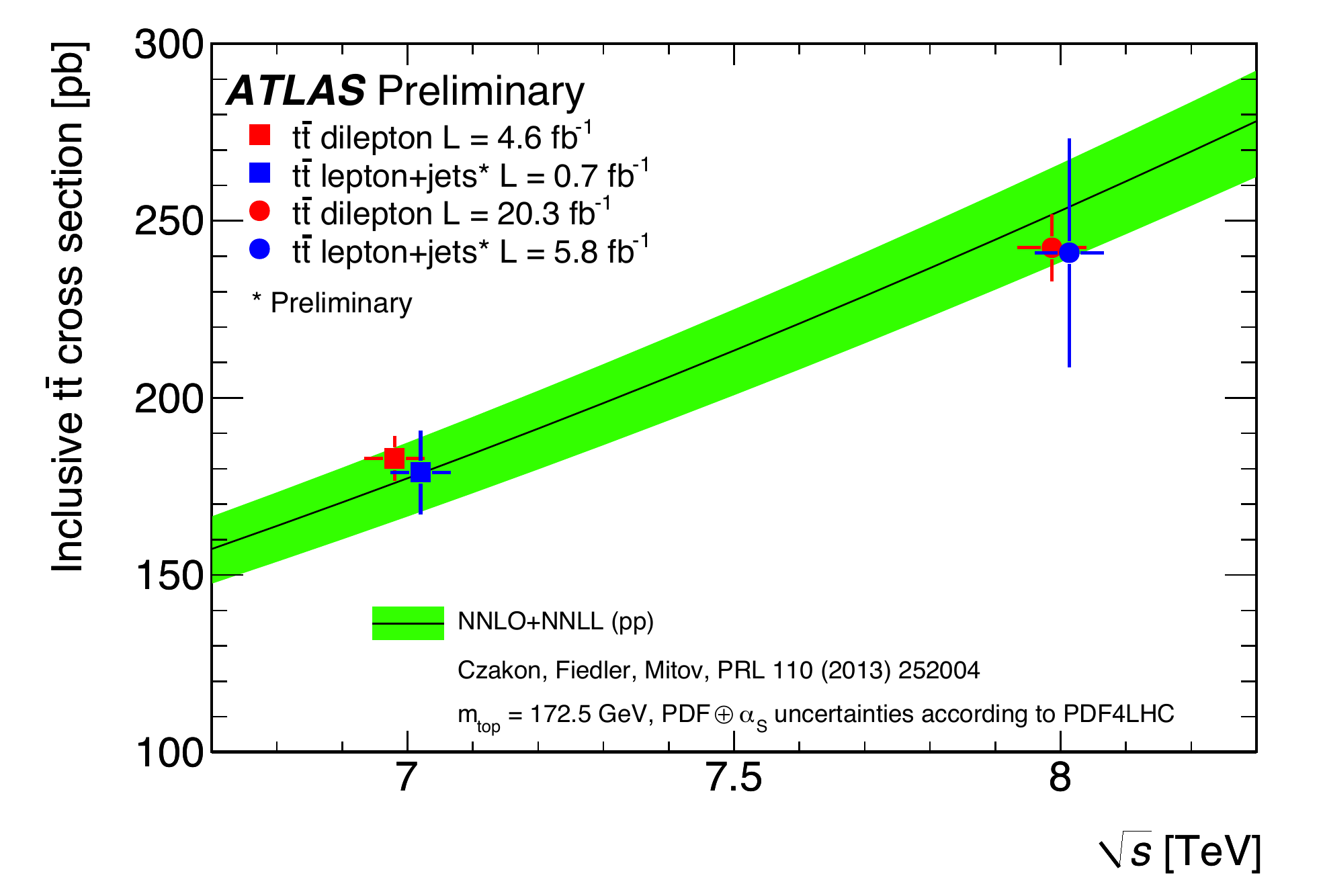}
        \caption{ATLAS top quark pair production cross section results at $\sqrt{s}=7$ TeV and $\sqrt{s}=8$ TeV.}
        \label{fig:atlasXS}
    \end{subfigure}
    ~ 
    \begin{subfigure}[t]{0.42\textwidth}
        \centering
        \includegraphics[width=0.95\textwidth]{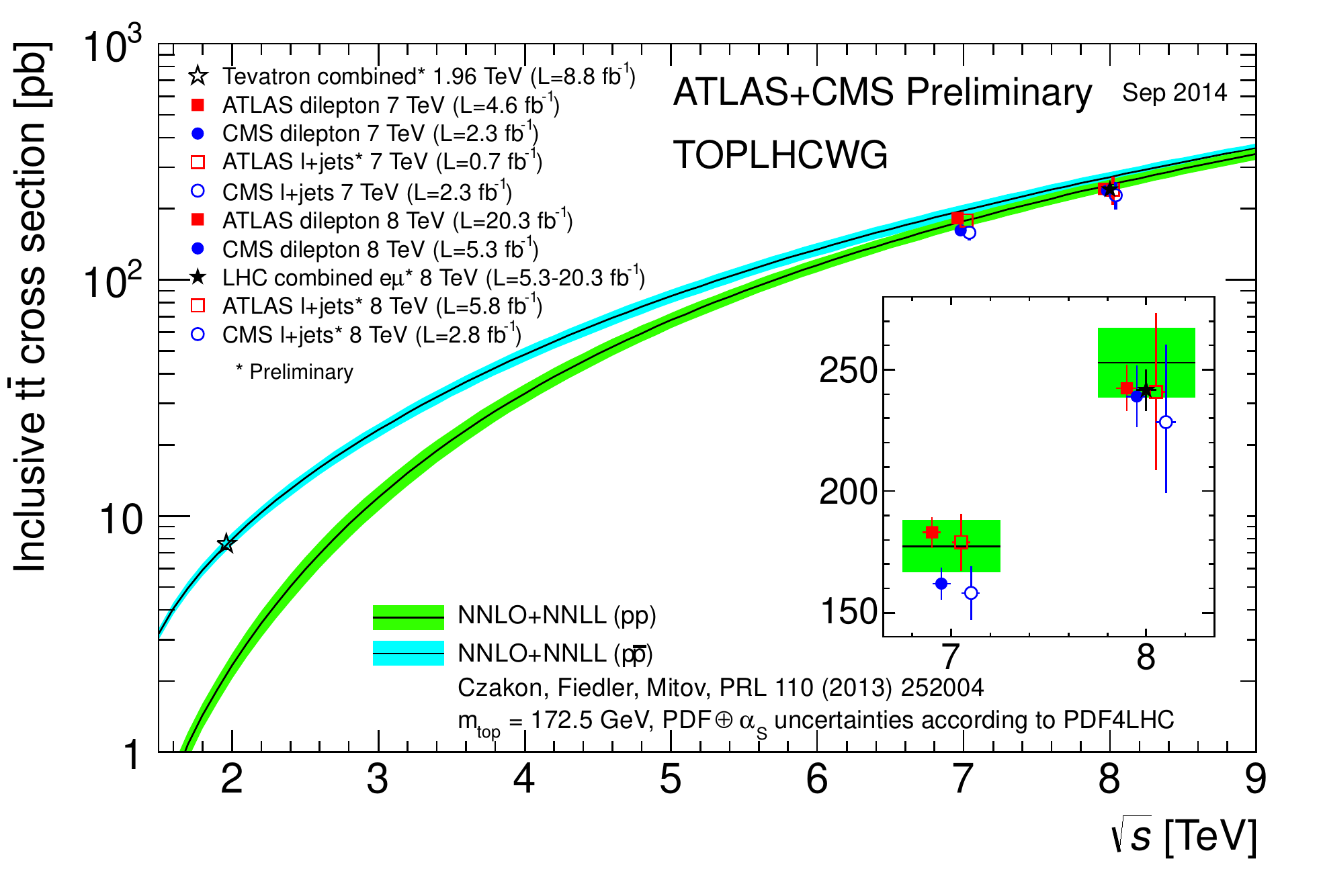}
        \caption{Global comparison of top quark pair production cross section results.}
        \label{fig:globalXS}
    \end{subfigure}        
    \caption{Summary of ATLAS top pair cross section results}
    \label{fig:CombinedXS}
\end{figure}

\vspace{-1.0em}
Figure~\ref{fig:atlasXS} shows the ATLAS $\sqrt{s}=7$ TeV and $\sqrt{s}=8$ TeV top quark pair production cross sections~\cite{ATLAS:TopSummary} compared to the theoretical predictions~\cite{Czakon:2013goa}, while Figure~\ref{fig:globalXS} includes the results from the Tevatron. Good agreement with theory is found globally.

\FloatBarrier

\end{document}

%% file: econfmacros.tex
\def\beq{\begin{equation}}
\def\eeq#1{\label{#1}\end{equation}}
\def\eeqn{\end{equation}}

\def\beqa{\begin{eqnarray}}
\def\eeqa#1{\label{#1}\end{eqnarray}}
\def\eeqan{\end{eqnarray}}

\let\bar=\overbar

\def\Dslash{\not{\hbox{\kern-4pt $D$}}}
\def\dslash{\not{\hbox{\kern-2pt $\del$}}}

\def\msb{{\bar{\ssstyle M \kern -1pt S}}}




%% file: jmorris.bbl
\begin{thebibliography}{99}

  

\bibitem{Aad:2008zzm} 
  ATLAS Collaboration,
  JINST {\bf 3}, S08003 (2008).
  
\bibitem{Evans:2008zzb} 
  L.~Evans and P.~Bryant (editors),
  JINST {\bf 3}, S08001 (2008).

\bibitem{Aad:2014zka} 
  ATLAS Collaboration,
  Phys.\ Rev.\ D {\bf 90}, 072004 (2014)
  [arXiv:1407.0371 [hep-ex]].
  
   
\bibitem{Cacciari:2008gp}
  M.~Cacciari, G.~P.~Salam and G.~Soyez,
  JHEP {\bf 0804}, 063 (2008)
  [arXiv:0802.1189 [hep-ph]].
  
\bibitem{Erdmann:2013rxa} 
  J.~Erdmann, S.~Guindon, K.~Kroeninger, B.~Lemmer, O.~Nackenhorst, A.~Quadt and P.~Stolte,
  Nucl.\ Instrum.\ Meth.\ A {\bf 748}, 18 (2014)
  [arXiv:1312.5595 [hep-ex]].
  
\bibitem{Hocker:1995kb} 
  A.~Hocker and V.~Kartvelishvili,
  Nucl.\ Instrum.\ Meth.\ A {\bf 372}, 469 (1996)
  [arXiv:hep-ph/9509307].
  
\bibitem{Group:2008vx} 
  R.~C.~Group {\it et al.} on behalf of the CDF Collaboration,
  [arXiv:0809.4670 [hep-ex]].
  
\bibitem{Corcella:2000bw} 
  G.~Corcella, I.~G.~Knowles, G.~Marchesini, S.~Moretti, K.~Odagiri, P.~Richardson, M.~H.~Seymour and B.~R.~Webber,
  JHEP {\bf 0101}, 010 (2001)
  [arXiv:hep-ph/0011363].
  
\bibitem{Mangano:2002ea} 
  M.~L.~Mangano, M.~Moretti, F.~Piccinini, R.~Pittau and A.~D.~Polosa,
  JHEP {\bf 0307}, 001 (2003)
  [arXiv:hep-ph/0206293].
  
\bibitem{Frixione:2002ik}
  S.~Frixione and B.~R.~Webber,
  JHEP {\bf 0206}, 029 (2002)
  [arXiv:hep-ph/0204244].
  
\bibitem{Nason:2004rx}
  P.~Nason,
  JHEP {\bf 0411 }, 040 (2004)
  [arXiv:hep-ph/0409146].
  
\bibitem{Frixione:2007vw}
  S.~Frixione, P.~Nason, C.~Oleari,
  JHEP {\bf 0711 }, 070 (2007)
  [arXiv:0709.2092 [hep-ph]]. 
  
\bibitem{Lai:2010vv} 
  H.~L.~Lai, M.~Guzzi, J.~Huston, Z.~Li, P.~M.~Nadolsky, J.~Pumplin and C.-P.~Yuan,
  Phys.\ Rev.\ D {\bf 82}, 074024 (2010)
  [arXiv:1007.2241 [hep-ph]].
  
\bibitem{Kidonakis:2010dk} 
  N.~Kidonakis,
  Phys.\ Rev.\ D {\bf 82}, 114030 (2010)
  [arXiv:1009.4935 [hep-ph]].
  
\bibitem{Martin:2009iq} 
  A.~D.~Martin, W.~J.~Stirling, R.~S.~Thorne and G.~Watt,
  Eur.\ Phys.\ J.\ C {\bf 63}, 189 (2009)
  [arXiv:0901.0002 [hep-ph]].
  
\bibitem{Ball:2012cx} 
  R.~D.~Ball, V.~Bertone, S.~Carrazza, C.~S.~Deans, L.~Del Debbio, S.~Forte, A.~Guffanti and N.~P.~Hartland {\it et al.},
  Nucl.\ Phys.\ B {\bf 867}, 244 (2013)
  [arXiv:1207.1303 [hep-ph]].
  
\bibitem{Aaron:2009aa} 
  H1 and ZEUS Collaborations, F.~D.~Aaron {\it et al.},
  JHEP {\bf 1001}, 109 (2010)
  [arXiv:0911.0884 [hep-ex]].
  
  
\bibitem{Aad:2014kva} 
  ATLAS Collaboration,
  [arXiv:1406.5375 [hep-ex]].
  
\bibitem{ATLAS:2012jyc} 
  ATLAS Collaboration,
  ATLAS-CONF-2012-149. https://cds.cern.ch/record/1493488
  
\bibitem{ATLAS:TopSummary} 
  ATLAS Collaboration,  
  https://atlas.web.cern.ch/Atlas/GROUPS/PHYSICS/CombinedSummaryPlots/TOP/  
 
\bibitem{Czakon:2013goa} 
  M.~Czakon, P.~Fiedler and A.~Mitov,
  Phys.\ Rev.\ Lett.\  {\bf 110}, 252004 (2013)
  [arXiv:1303.6254 [hep-ph]].



\end{thebibliography}
